\documentclass[10pt]{article}

\usepackage{microtype}
\usepackage{graphicx}
\usepackage{subcaption}
\usepackage{booktabs} 

\usepackage{hyperref}

\usepackage[accepted]{icml2026}

\usepackage{amsmath}
\usepackage{amssymb}
\usepackage{mathtools}
\usepackage{amsthm}
\usepackage{enumitem}
\usepackage{makecell}

\usepackage[capitalize,noabbrev]{cleveref}

\theoremstyle{plain}

\theoremstyle{definition}

\theoremstyle{remark}

\usepackage[textsize=tiny]{todonotes}

\icmltitlerunning{Set-Aggregated Genome Embeddings for Microbiome Abundance Prediction}

\begin{document}
	
	\twocolumn[
	\icmltitle{Set-Aggregated Genome Embeddings for Microbiome Abundance Prediction}
	
	
	
	\icmlsetsymbol{equal}{*}
	
	\begin{icmlauthorlist}
		\icmlauthor{Younhun Kim}{bwh,hms}
		\icmlauthor{Georg K. Gerber}{bwh,hms}
		\icmlauthor{Travis E. Gibson}{bwh,hms}
	\end{icmlauthorlist}
	
	\icmlaffiliation{bwh}{Brigham and Women's Hospital, Boston, MA, USA}
	\icmlaffiliation{hms}{Harvard Medical School, Boston, MA, USA}
	
	\icmlcorrespondingauthor{Travis E. Gibson}{tegibson@bwh.harvard.edu}
	
	\icmlkeywords{Machine Learning, ICML, Biology, Bioinformatics, Genomics, Genomic Language Model, Large Language Model, Microbiome}
	
	\vskip 0.3in
	]
	
	
	
	\printAffiliationsAndNotice{}  
	
	\begin{abstract}
		Microbiome functions are encoded within the genes of the community-wide metagenome.
		A natural question is whether properties of a microbial community can be predicted just from knowing the raw DNA sequences of its members.
		In this work, we employ set-aggregated genome embeddings (SAGE) to predict community-level abundance profiles, exploiting the few-shot learning capabilities of genomic language models (GLMs). 
		We benchmark this approach to show improved generalization on novel genomes compared to classical bioinformatics approaches. Model ablation shows that community-level latent representations directly result in improved performance. Lastly, we demonstrate the benefits of intermediate transformations between latent representations and demonstrate the differences between GLM embedding choices.
	\end{abstract}
	
	\section{Introduction}
	
	\begin{figure}[t]
		\begin{center}
			\centerline{\includegraphics[width=0.92\columnwidth]{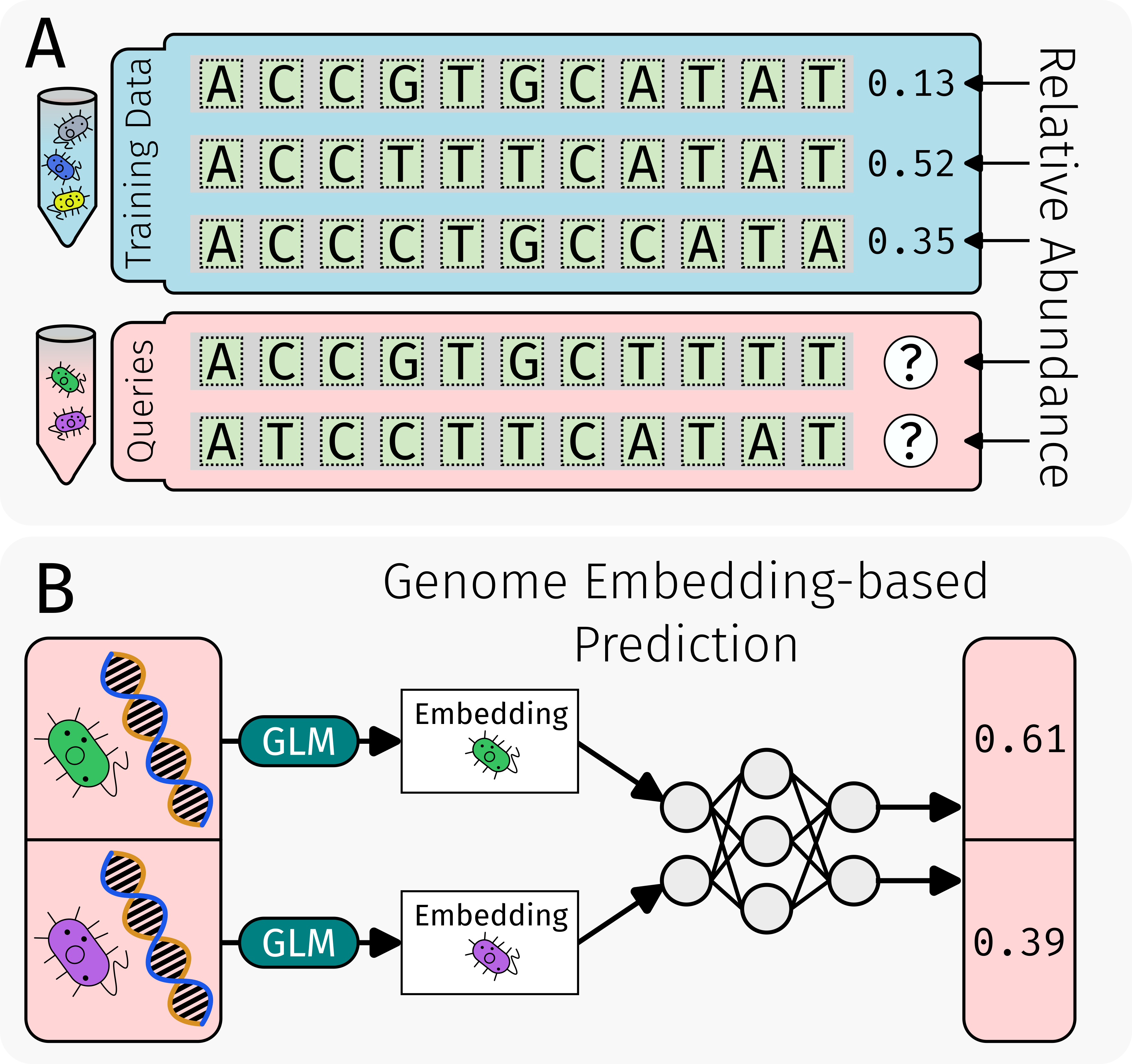}}
			\caption{
				Overview of (a) the predictive task and (b) the deep-learning approach. The primary goal is to use genomic sequences directly to produce abundance predictions per microbiome sample.
			}
			\label{fig:overview}
		\end{center}
	\end{figure}
	
	Microbiomes are intensely studied for their wide array of impactful applications, notably in the area of human health and medicine \cite{gilbert2018current,fan2021gut}. 
	A big central question shared by all major applications is: \emph{can we predict properties of never-before-seen microbiomes?} That is, given knowledge of what taxa are present, can we make predictions about a target desired property? 
	This is spurred by a big hurdle in microbiome science, namely that it is difficult to isolate \& culture arbitrary bacterial strains at scale. Thus, it is infeasible to understand all novel microbiomes by growing or isolating each individual member \citep{lagier2018culturing}.
	
	Thanks to advancements in molecular sequencing, it is now commonplace to perform DNA sequencing of microbiomes, producing  reads from the ``pool'' of genomes (\emph{metagenome}) present in any sample. 
	Very few algorithms utilize genomic sequence information end-to-end.
	Recent line of research has showed that genomic language models (GLMs) such as Evo \cite{evo} and DNABERT \cite{dnabert} can be feature-extracted for microbiome tasks. The two prototypical tasks which have appeared in literature are environmental source prediction and host phenotype prediction \cite{setbert2025lugwig,yoo2025abundance}. 
	
	However, many present-day questions in the microbiome field involve predictions on a per-species or per-strain basis, one for each found in the input community.
	Examples include bacterial interactions \cite{mdsine2}, perturbational response \cite{ng2019recovery}, and engraftment prediction after fecal microbiota transplant \cite{smillie2018strain}.
	It is also unclear whether the community-level pooling representations, as seen in Set-Attention \cite{lee2019set}, is helpful for encoding microbiomes, if it can be improved, or would benefit from different choices of GLMs.

	To understand the above questions, we investigate the task of \emph{relative abundance prediction} directly from genomic sequences (Figure~\ref{fig:overview}a). 
	Here, we assume that the raw sequencing reads from each microbiome experiment have been pre-processed into clusters with representative sequences (``taxa'').
	Our datasets come with a ``ground truth'' table of each dataset; each $(i, j)$ cell is the observed abundance of taxa $j$ in microbiome sample $i$, normalized so each row sums to $1$.
	The target objective is to predict the abundance table (each $(i, j)$ cell) only using information about which taxa $j$ are found per sample (the support of each row) and the representative gene sequence(s) of each $j$. The predictive models we study are neural networks which take the nucleotide sequence embeddings as input (Figure~\ref{fig:overview}b), with a particular focus on a family of models that use permutation-invariant pooling on to latently represent each taxa and community.

	\section{Datasets}
	
	We used two datasets to train and evaluate our models. The first is the American Gut Project (AGP) \cite{mcdonald2018americangut}, a 16S-V4 amplicon sequencing dataset processed via the Human Microbiome Compendium \cite{humanmicrobiome2007} using DADA2 \cite{dada2}. This yields taxa-specific gene sequences referred to as Amplicon Sequence Variants (ASVs). The second is the whole-metagenome sequencing (WMS) dataset from the MetaPhlAn4 publication \cite{metaphlan4}, where taxa are Species-level Genome Bins (SGBs). 
	ASVs are \emph{de novo} sequences; SGBs are reference-database clusters defined by marker genes. After quality screening (Appendix~\ref{sec:data-qc}), the 16S dataset contained 3,851 healthy adult samples and the WMS dataset contained 6,015 healthy adult samples. We train separate models per dataset.

	\subsection{Train-Test Split}
	We benchmark using two main types of splitting methodologies.
	The first is a simple, uniformly random train-test split (``random split'') where samples are globally shuffled and split at a 4:1 train/test ratio.
	Second, to test generalization to unseen microbiome compositions, we constructed an out-of-distribution test split  (``JS-maximizing split''). To be precise, we computed pairwise Jensen-Shannon (JS) divergence between samples and applied PCoA \cite{pcoa1996gower} to the resulting distance matrix. The first principal coordinate was used to split samples at a 4:1 train/test ratio, placing more compositionally distant samples in the test set.
	

	\section{Predictive Models}
	
	Here, we define methods sorted into two categories: 
	(1) distance-aware, sequence-naive (methods can only use distances computed by bioinformatics sequence comparison), and
	(2) sequence-aware embedding methods.
	\#2 is the main focus of our paper; \#1 serves as a baseline.
	
	
	
	\subsection{Nearest-Neighbor Baseline Methods}
	We assume knowledge of a pairwise distance metric $\delta$ between taxa, spanning both the training-set and test-set taxa.
	For 16S, $\delta$ is derived from \% sequence identity via Mothur \cite{mothur2009}; for WMS, $\delta$ is the phylogenetic tree distance from PhyloPhlAn3 \cite{phylophlan3}. 
	Each outputs a relative abundance probability vector:
	\begin{enumerate}[noitemsep,topsep=0pt]
		\item \emph{Uniform}: Assigns $1/N$ relative abundance to each taxon.
		\item \emph{Global Average}: maps each query taxon to its nearest neighbor in the training set; weights by geometric-mean abundance across all training samples.
		\item \emph{$k$NN Average}: same nearest-neighbor mapping of taxa, but weights by geometric-mean abundance within the $k$ most similar training samples by Jaccard similarity.
	\end{enumerate}
	These methods were previously benchmarked in \cite{bashan2024predict} and shown to outperform sequence and distance-naive neural network. Distance-aware neural networks were not tested.

	\subsection{Phylogenetic Embedding Methods}
	We include distance-aware, sequence-naive neural network methods by first constructing UMAP embeddings \cite{umap2018mcinnes} from $\delta$, sweeping output dimensions $d \in \{20,40,60,80,100\}$.
	Each taxon is represented as a  $1 \times d$ vector and passed into a deep-learning model (Section~\ref{sec:model-architecture}) specific to the choice of $d$.
	
	%
	
	\subsection{Genomic Sequence Embeddings}
	This family of sequence-aware neural networks is the main focus of our paper.
	We compare three pre-trained genomic language models (GLMs): Evo \cite{evo}, Evo 2 \cite{evo2}, and DNABERT-S \cite{dnabert-s}.
	These models were selected to represent the current landscape of microbe-focused GLM and restricted to those explicitly tested on microbial genome prediction, as opposed to GLMs primarily trained \& solely evaluated on human sequences.
	For Evo/Evo 2, we extract intermediate layer representations ($d=4096$); for DNABERT-S we use the final layer ($d=768$). Weights were not fine-tuned. 
	Each 16S ASV is embedded as a single $d$-dimensional vector; each WMS SGB is embedded across its $g$ PhyloPhlAn marker genes, producing a $g \times d$ matrix. For WMS, we apply incremental PCA \cite{ross2008incremental} across the entire reference database gene set (independent from any train-test split) to reduce dimensionality and manage memory constraints.

	\section{Deep Learning Architecture \& Training}
	\label{sec:model-architecture}
	
	\begin{figure}[t]
		\begin{center}
			\centerline{\includegraphics[width=0.90\columnwidth]{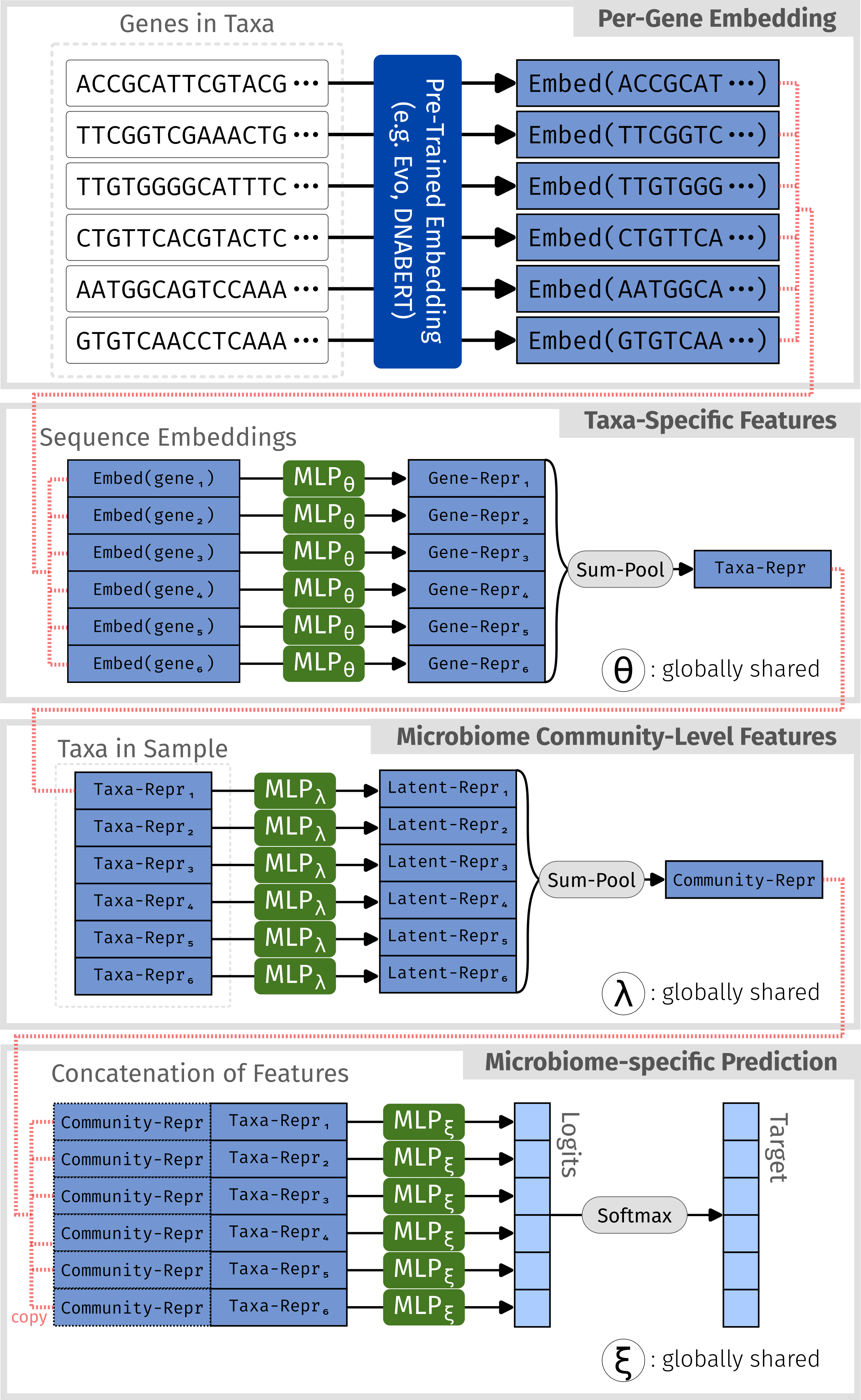}}
			\caption{
				Architectural diagram of the benchmarked SAGE implementation.
				The architecture is divided into modules implementing successive hierarchies: microbiomes are sets of taxa, taxa are sets of genes, and genes are represented by sequences.
				Intermediate MLPs are broadcasted across each slice, to serve as latent transformations between represented spaces.
			}
			\label{fig:architecture}
		\end{center}
	\end{figure}
	
	Each input taxa is generically a $g \times d$ matrix, representing $g$ genes embedded using a $d$-dimensional GLM output. Thus, a microbiome sample consisting of $t$ taxa is a (padded and masked) $t \times g \times d$ tensor.
	To handle these types of input representations, we dubbed the architectural strategy \emph{Set-Aggregated Genomic Embedding} (SAGE), to refer to the pooling of genomic features in each taxa. 
	
	The architecture (Figure~\ref{fig:architecture}) can be understood in terms of its \emph{modules} (gray boxes).
	Each ``level'' of aggregation is done by first transforming each input representation using a multi-layer perceptron (MLP) submodule, which is broadcasted across each of the constituent feature vector slices. Then, the transformed latent representations are sum-pooled.
	This is a canonical example of a ``permutation-equivariant'' neural network \cite{lee2019set}, which has the property of being agnostic to the ordering of the microbial taxa, and to the ordering of the genes.
	The final outputs are taxa-specific logits, which is produced via a third MLP submodule that uses the concatenation of the taxa-specific and community-wide features.
	
	Alternatively, we could have hierarhically composed Set-Attention blocks \cite{lee2019set} as seen in \cite{setbert2025lugwig,yoo2025abundance}, instead of sum-pooling. Our ablation analysis (Figure~\ref{fig:ablation_mlp}) shows that performance saturates on our current architecture, suggesting that additional complexity and capacity may not meaningfully improve generalization on our datasets. However, an architecture like Set-Attention is quite attractive for future work, as the inductive bias better represents interactions between taxa found in dynamical systems models \cite{mdsine2,thompson2026physics}.
	
	
	We trained all SAGE models using the Adam optimizer \cite{kingma2014adam} on the Kullback-Liebler (KL) divergence loss, initialized with $10^{-4}$ learning rate and a cosine annealing schedule. Lastly, we did not use the test data in any way during training, such as early stopping, to avoid data leakage. Instead, all models were trained for exactly 80 epochs, which appeared sufficient to reach a local minimum of the training loss in all scenarios.
	
	\section{Evaluation Results}
	
	\begin{figure}[t]
		\begin{center}
			\textsc{16S ASV Dataset}
			\centerline{\includegraphics[width=\columnwidth]{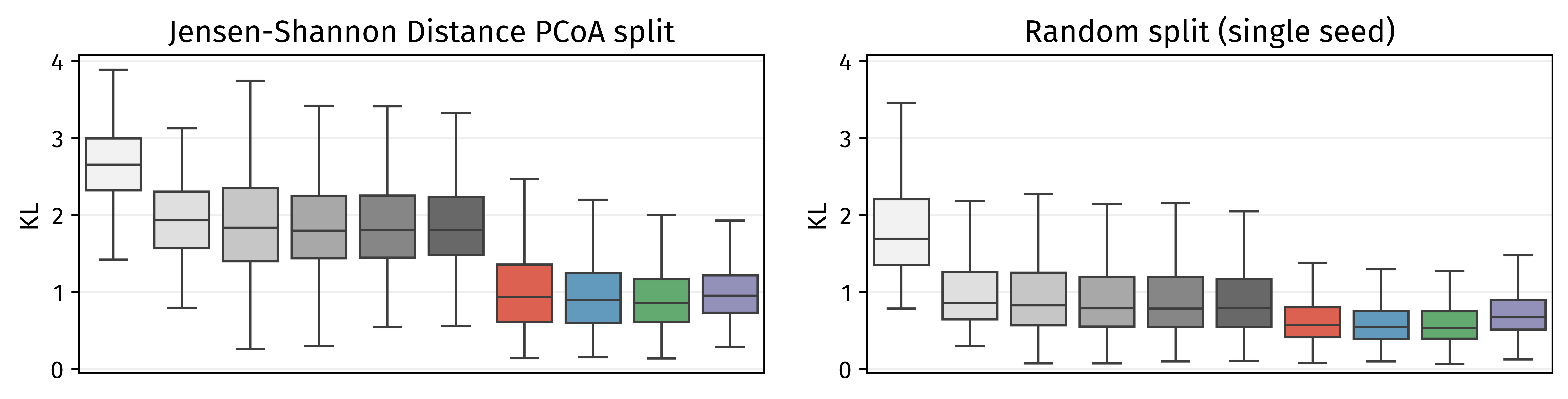}}
			\textsc{WMS SGB Dataset}
			\centerline{\includegraphics[width=\columnwidth]{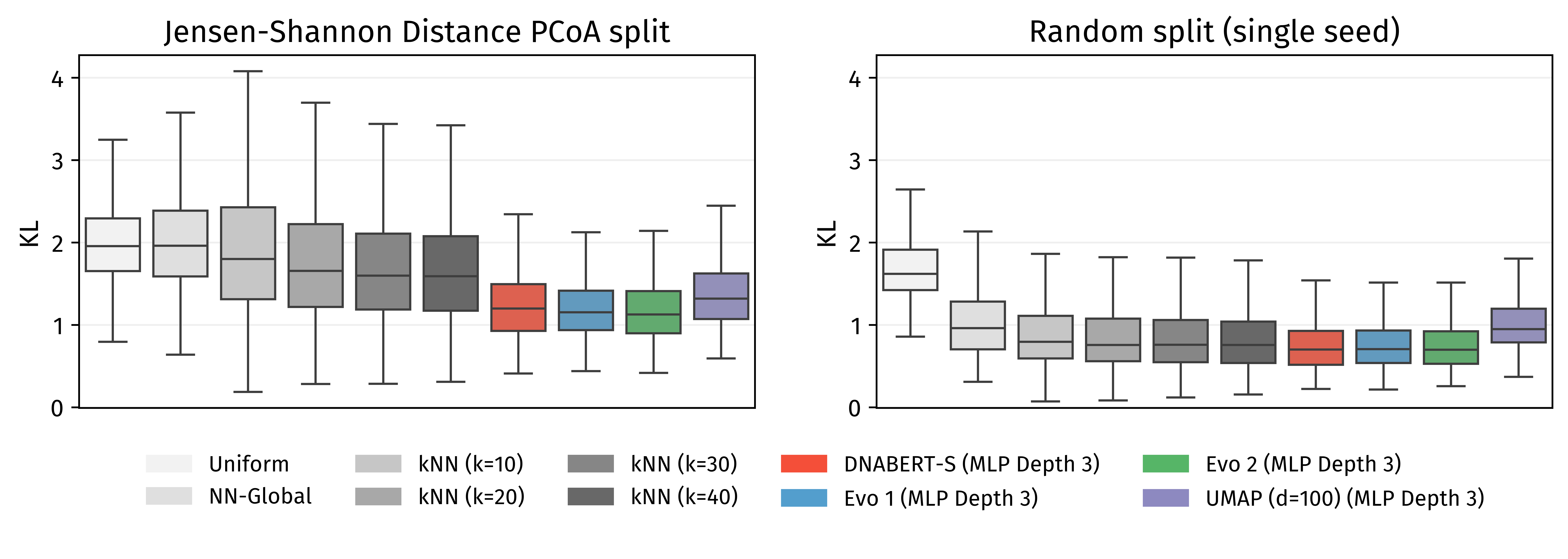}}
			\caption{
				Benchmarking results on the 16S ASV dataset (top row) and the WMS SGB dataset (bottom row).
				We 
			}
			\label{fig:benchmark}
		\end{center}
	\end{figure}
	
	\subsection{\emph{De Novo} Sequence, Single-Gene Evaluation (16S ASV)}
	
	We evaluate test-set KL divergence as our primary error metric (Figure~\ref{fig:benchmark}, top row; median loss values in Table~\ref{table:benchmark_asv}) on the 16S dataset. 
	A quick sanity check shows that the trend from \cite{bashan2024predict} is recovered, where $k$NN outperforms Global Averaging.
	On the JS-maximizing split, all deep-learning methods consistently outperform all baselines ($p$-values in Tables~\ref{table:pval_pcoa_asv},\ref{table:pval_rng_asv}).
	Apparent differences between GLM choices can largely be attributed to embedding sizes and parameter counts (Section~\ref{sec:model-ablation}). Varying MLP depth generally does not make a difference, except for random splits on UMAP embeddings (Figure~\ref{fig:mlp_depths}(a,b)).
	
	Across random splits, all methods improve with sequence-based models reaching KL $\approx 0.55$. 
	Phylogenetic embedding models matched sequence models closely in KL. This is true across all tested UMAP output dimensions $d$, as model performance increases with $d$ but with diminishing returns and only on randomized splits (Figure~\ref{fig:mlp_depths}(c,d)).
	
	\subsection{Reference Sequence, Multi-Gene Evaluation (WGS SGB)}
	The WMS benchmark (Figure~\ref{fig:benchmark}, bottom row; median loss values in Table~\ref{table:benchmark_sgb}) showed similar patterns, although there was a smaller overall performance gap between the averaging methods and the neural networks.
	Still, the sequence-embedding models outperformed baselines in KL  ($p$-values in Tables~\ref{table:pval_pcoa_sgb},\ref{table:pval_rng_sgb}). 
	Adding more marker genes (PhyloPhlAn + MetaPhlAn) did not improve performance (Figure~\ref{fig:metaphlan_gene_set}), suggesting saturation with the first gene set alone; a full gene ablation analysis is left to future work.
	
	\subsection{Model Ablation} \label{sec:model-ablation}
	
	\begin{figure}[h]
		\begin{center}
			\centerline{\includegraphics[width=\columnwidth]{asv_ablation.png}}
			\caption{
				Benchmark of SAGE models for ablating community-level pooling. Models are either (a) given zero vectors for community-level representations during evaluation, or (b) are fully re-trained with community representations removed.
			}
			\label{fig:ablation_pool}
		\end{center}
	\end{figure}
	
	To verify the effectiveness of including community representations, we conducted an ablation study on the community-level pooling module (Figure~\ref{fig:ablation_pool}), using the JS-maximizing split 16S dataset. 
	Models were either given zero-vectors for community representations during evaluation (Figure~\ref{fig:ablation_pool}, left), or were re-trained with the representations removed outright (Figure~\ref{fig:ablation_pool}, right).
	The ablated intermediary representations resulted in worse performance in either setting, across all GLM embedding types.
	
	\begin{figure}[h]
		\begin{center}
			\centerline{\includegraphics[width=\columnwidth]{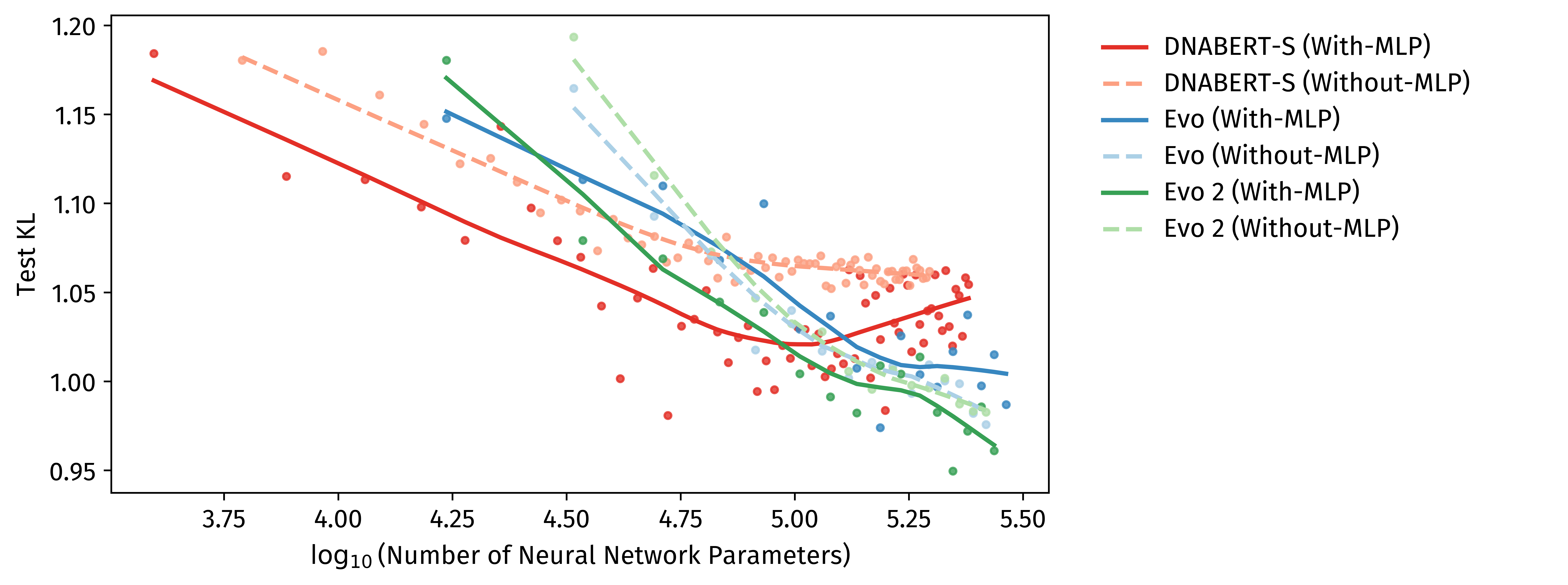}}
			\caption{
				MLP and model capacity ablation analysis.
				Models are trained with and without the MLP transformations in the taxa-pooling and community-pooling modules, sweeping across parameter counts. Curves are local fits (LOWESS).
			}
			\label{fig:ablation_mlp}
		\end{center}
	\end{figure}
	
	Lastly, we explored ablation of $\text{MLP}_\theta, \text{MLP}_\lambda$ from Figure~\ref{fig:ablation_mlp}.
	A priori, their usefulness is not obvious: a previous work \cite{yoo2025abundance} showed that pooling without intermediate transformations results in worse performance, but only marginally and selectively depending on task.
	Our analysis found that the ``without-MLP'' architectures either had poorer (DNABERT-S variants) or approximately equal (Evo, Evo 2) test-loss scaling as parameter count was increased.
	This analysis also showed that DNABERT-S results in SAGE performance quickly worsening (due to overfitting and not undertraining), whereas Evo 2 results in SAGE performance continuing to improve well past $10^5$ parameters.

	\section{Discussion}
	
	Our results support the main hypothesis: genome embedding aggregations results in better generalization to novel sequences unseen during training. 
	Still, the WMS dataset's smaller performance gap was curious to us.
	We speculate that this is due to the dataset's nature where species-level reference sequences are used, rather than \emph{de novo} variants. 
	
	The ablation findings confirm that including extra layers produce more efficiently trainable models. 
	Our dataset is likely too small to extrapolate scaling behavior in the manner of \cite{kaplan2020scaling,hoffmann2022training}.
	Rather, this analysis characterizes the sensitivity of performance to model capacity within our training regime.
	Indeed, difference in performance is only visible given a limited compute budget (e.g. memory capacity) and limited dataset size. While the former is solvable with a larger budget, the latter is not easily fixable. Microbiome sample collection is difficult to scale, particularly in human studies with a huge diversity of both host and microbiome.
	Our observed test KL loss of $0.8 \sim 0.9$ may seem high, but the reader should note that we did not use any of the host features in the predictions. With a larger, cleaner, and more well-annotated dataset, it may be possible to push this performance even further.
	
	In future work, we hope to explore other types of inductive biases depending on microbiome-specific tasks. We are especially excited to explore the potential of longer \emph{de novo} long-read sequencing using this approach, which would provide the ``best of both worlds'' of the single-gene and whole-genome datasets.
	
	

	
	\newpage
	\bibliography{manuscript}

@article{fan2021gut,
	title={Gut microbiota in human metabolic health and disease},
	author={Fan, Yong and Pedersen, Oluf},
	journal={Nature Reviews Microbiology},
	volume={19},
	number={1},
	pages={55--71},
	year={2021},
	publisher={Nature Publishing Group UK London}
}

@article{gilbert2018current,
	title={Current understanding of the human microbiome},
	author={Gilbert, Jack A and Blaser, Martin J and Caporaso, J Gregory and Jansson, Janet K and Lynch, Susan V and Knight, Rob},
	journal={Nature medicine},
	volume={24},
	number={4},
	pages={392--400},
	year={2018},
	publisher={Springer New York New York}
}

@article{lagier2018culturing,
	title={Culturing the human microbiota and culturomics},
	author={Lagier, Jean-Christophe and Dubourg, Gr{\'e}gory and Million, Matthieu and Cadoret, Fr{\'e}d{\'e}ric and Bilen, Melhem and Fenollar, Florence and Levasseur, Anthony and Rolain, Jean-Marc and Fournier, Pierre-Edouard and Raoult, Didier},
	journal={Nature Reviews Microbiology},
	volume={16},
	number={9},
	pages={540--550},
	year={2018},
	publisher={Nature Publishing Group UK London}
}

@article{ng2019recovery,
	title={Recovery of the gut microbiota after antibiotics depends on host diet, community context, and environmental reservoirs},
	author={Ng, Katharine Michelle and Aranda-D{\'\i}az, Andr{\'e}s and Tropini, Carolina and Frankel, Matthew Ryan and Van Treuren, William and O’Loughlin, Colleen T and Merrill, Bryan Douglas and Yu, Feiqiao Brian and Pruss, Kali M and Oliveira, Rita Almeida and others},
	journal={Cell host \& microbe},
	volume={26},
	number={5},
	pages={650--665},
	year={2019},
	publisher={Elsevier}
}

@article{metaphlan4,
	title={Extending and improving metagenomic taxonomic profiling with uncharacterized species using MetaPhlAn 4},
	author={Blanco-M{\'\i}guez, Aitor and Beghini, Francesco and Cumbo, Fabio and McIver, Lauren J and Thompson, Kelsey N and Zolfo, Moreno and Manghi, Paolo and Dubois, Leonard and Huang, Kun D and Thomas, Andrew Maltez and others},
	journal={Nature biotechnology},
	volume={41},
	number={11},
	pages={1633--1644},
	year={2023},
	publisher={Nature Publishing Group US New York}
}

@article{evo,
	title={Sequence modeling and design from molecular to genome scale with Evo},
	author={Nguyen, Eric and Poli, Michael and Durrant, Matthew G and Kang, Brian and Katrekar, Dhruva and Li, David B and Bartie, Liam J and Thomas, Armin W and King, Samuel H and Brixi, Garyk and others},
	journal={Science},
	volume={386},
	number={6723},
	pages={eado9336},
	year={2024},
	publisher={American Association for the Advancement of Science}
}

@article{dnabert-s,
	title={Dnabert-s: Learning species-aware dna embedding with genome foundation models},
	author={Zhou, Zhihan and Wu, Weimin and Ho, Harrison and Wang, Jiayi and Shi, Lizhen and Davuluri, Ramana V and Wang, Zhong and Liu, Han},
	journal={arXiv preprint arXiv:2402.08777},
	volume={10},
	year={2024}
}

@article{dnabert,
	title={DNABERT: pre-trained Bidirectional Encoder Representations from Transformers model for DNA-language in genome},
	author={Ji, Yanrong and Zhou, Zhihan and Liu, Han and Davuluri, Ramana V},
	journal={Bioinformatics},
	volume={37},
	number={15},
	pages={2112--2120},
	year={2021},
	publisher={Oxford University Press}
}

@article{mdsine2,
	title={Learning ecosystem-scale dynamics from microbiome data with MDSINE2},
	author={Gibson, Travis E and Kim, Younhun and Acharya, Sawal and Kaplan, David E and DiBenedetto, Nicholas and Lavin, Richard and Berger, Bonnie and Allegretti, Jessica R and Bry, Lynn and Gerber, Georg K},
	journal={Nature Microbiology},
	volume={10},
	number={10},
	pages={2550--2564},
	year={2025},
	publisher={Nature Publishing Group UK London}
}

@article{thompson2026physics,
	title={Physics-constrained neural ordinary differential equation models to discover and predict microbial community dynamics},
	author={Thompson, Jaron and Connors, Bryce M and Zavala, Victor M and Venturelli, Ophelia S},
	journal={Proceedings of the National Academy of Sciences},
	volume={123},
	number={13},
	pages={e2517661123},
	year={2026},
	publisher={National Academy of Sciences}
}

@article{bashan2024predict,
	title={Model-free prediction of microbiome compositions},
	author={Asher, Eitan E and Bashan, Amir},
	journal={Microbiome},
	volume={12},
	number={1},
	pages={17},
	year={2024},
	publisher={Springer}
}

@article{evo2,
	title={Genome modeling and design across all domains of life with Evo 2},
	author={Brixi, Garyk and Durrant, Matthew G and Ku, Jerome and Poli, Michael and Brockman, Greg and Chang, Daniel and Gonzalez, Gabriel A and King, Samuel H and Li, David B and Merchant, Aditi T and others},
	journal={BioRxiv},
	pages={2025--02},
	year={2025},
	publisher={Cold Spring Harbor Laboratory}
}

@article{phylophlan3,
	title={Precise phylogenetic analysis of microbial isolates and genomes from metagenomes using PhyloPhlAn 3.0},
	author={Asnicar, Francesco and Thomas, Andrew Maltez and Beghini, Francesco and Mengoni, Claudia and Manara, Serena and Manghi, Paolo and Zhu, Qiyun and Bolzan, Mattia and Cumbo, Fabio and May, Uyen and others},
	journal={Nature communications},
	volume={11},
	number={1},
	pages={2500},
	year={2020},
	publisher={Nature Publishing Group UK London}
}

@article{humanmicrobiome2007,
	title={The human microbiome project},
	author={Turnbaugh, Peter J and Ley, Ruth E and Hamady, Micah and Fraser-Liggett, Claire M and Knight, Rob and Gordon, Jeffrey I},
	journal={Nature},
	volume={449},
	number={7164},
	pages={804--810},
	year={2007},
	publisher={Nature Publishing Group UK London}
}

@article{setbert2025lugwig,
	title={SetBERT: the deep learning platform for contextualized embeddings and explainable predictions from high-throughput sequencing},
	author={Ludwig, David W and Guptil, Christopher and Alexander, Nicholas R and Zhalnina, Kateryna and Wipf, Edi M-L and Khasanova, Albina and Barber, Nicholas A and Swingley, Wesley and Walker, Donald M and Phillips, Joshua L},
	journal={Bioinformatics},
	volume={41},
	number={7},
	pages={btaf370},
	year={2025},
	publisher={Oxford University Press}
}

@article{yoo2025abundance,
	title={Abundance-Aware Set Transformer for Microbiome Sample Embedding},
	author={Yoo, Hyunwoo and Rosen, Gail},
	journal={arXiv preprint arXiv:2508.11075},
	year={2025}
}

@article{mcdonald2018americangut,
	title={American gut: an open platform for citizen science microbiome research},
	author={McDonald, Daniel and Hyde, Embriette and Debelius, Justine W and Morton, James T and Gonzalez, Antonio and Ackermann, Gail and Aksenov, Alexander A and Behsaz, Bahar and Brennan, Caitriona and Chen, Yingfeng and others},
	journal={Msystems},
	volume={3},
	number={3},
	pages={10--1128},
	year={2018},
	publisher={American Society for Microbiology 1752 N St., NW, Washington, DC}
}

@article{ross2008incremental,
	title={Incremental learning for robust visual tracking},
	author={Ross, David A and Lim, Jongwoo and Lin, Ruei-Sung and Yang, Ming-Hsuan},
	journal={International journal of computer vision},
	volume={77},
	number={1},
	pages={125--141},
	year={2008},
	publisher={Springer}
}

@article{pcoa1996gower,
	title={Some distance properties of latent root and vector methods used in multivariate analysis},
	author={Gower, John C},
	journal={Biometrika},
	volume={53},
	number={3-4},
	pages={325--338},
	year={1966},
	publisher={Oxford University Press}
}

@article{umap2018mcinnes,
	title={Umap: Uniform manifold approximation and projection for dimension reduction},
	author={McInnes, Leland and Healy, John and Melville, James},
	journal={arXiv preprint arXiv:1802.03426},
	year={2018}
}

@article{mothur2009,
	title={Introducing mothur: open-source, platform-independent, community-supported software for describing and comparing microbial communities},
	author={Schloss, Patrick D and Westcott, Sarah L and Ryabin, Thomas and Hall, Justine R and Hartmann, Martin and Hollister, Emily B and Lesniewski, Ryan A and Oakley, Brian B and Parks, Donovan H and Robinson, Courtney J and others},
	journal={Applied and environmental microbiology},
	volume={75},
	number={23},
	pages={7537--7541},
	year={2009},
	publisher={American Society for Microbiology}
}

@article{dada2,
	title={DADA2: High-resolution sample inference from Illumina amplicon data},
	author={Callahan, Benjamin J and McMurdie, Paul J and Rosen, Michael J and Han, Andrew W and Johnson, Amy Jo A and Holmes, Susan P},
	journal={Nature methods},
	volume={13},
	number={7},
	pages={581--583},
	year={2016},
	publisher={Nature Publishing Group US New York}
}

@inproceedings{lee2019set,
	title={Set transformer: A framework for attention-based permutation-invariant neural networks},
	author={Lee, Juho and Lee, Yoonho and Kim, Jungtaek and Kosiorek, Adam and Choi, Seungjin and Teh, Yee Whye},
	booktitle={International conference on machine learning},
	pages={3744--3753},
	year={2019},
	organization={PMLR}
}

@article{kaplan2020scaling,
	title={Scaling laws for neural language models},
	author={Kaplan, Jared and McCandlish, Sam and Henighan, Tom and Brown, Tom B and Chess, Benjamin and Child, Rewon and Gray, Scott and Radford, Alec and Wu, Jeffrey and Amodei, Dario},
	journal={arXiv preprint arXiv:2001.08361},
	year={2020}
}

@article{hoffmann2022training,
	title={Training compute-optimal large language models},
	author={Hoffmann, Jordan and Borgeaud, Sebastian and Mensch, Arthur and Buchatskaya, Elena and Cai, Trevor and Rutherford, Eliza and Casas, DDL and Hendricks, Lisa Anne and Welbl, Johannes and Clark, Aidan and others},
	journal={arXiv preprint arXiv:2203.15556},
	volume={10},
	year={2022}
}

@article{kingma2014adam,
	title={Adam: A method for stochastic optimization},
	author={Kingma, Diederik P and Ba, Jimmy},
	journal={arXiv preprint arXiv:1412.6980},
	year={2014}
}

@article{smillie2018strain,
	title={Strain tracking reveals the determinants of bacterial engraftment in the human gut following fecal microbiota transplantation},
	author={Smillie, Christopher S and Sauk, Jenny and Gevers, Dirk and Friedman, Jonathan and Sung, Jaeyun and Youngster, Ilan and Hohmann, Elizabeth L and Staley, Christopher and Khoruts, Alexander and Sadowsky, Michael J and others},
	journal={Cell host \& microbe},
	volume={23},
	number={2},
	pages={229--240},
	year={2018},
	publisher={Elsevier}
}
	\bibliographystyle{icml2026}
	
	\newpage
	\appendix
	\onecolumn
	\setcounter{figure}{0}
	\setcounter{table}{0}
	\renewcommand{\thefigure}{S\arabic{figure}}
	\renewcommand{\thetable}{S\arabic{table}}
	\section{Data Processing \& Quality Control}
	\label{sec:data-qc}
	
	\subsection{American Gut Project (16S Amplicon)}
	The AGP dataset sequenced amplicons of a short $\sim$125bp hypervariable segment (the v4 region) of the 16S gene.
	We first downloaded the DADA2-processed abundance tables and ASV sequence FASTA files from the Human Microbiome Compendium.
	To remove outliers which might cause unnecessarily confusing technical bias, we initially performed a 2-step filtering on this dataset.
	First, to reduce the effect of extremal deviations in molecular amplification and/or sequencing, we removed samples with overall read count in the top 2.5\% and bottom 2.5\%. 
	Next, to eliminate ASVs and samples with the highest likelihood of having anomalous DADA2 outputs, we removed all ASVs that do not exceed a read count of 10 in at least one sample, and then removed all samples that either have too few ASVs ($\leq 50$) or too many ASVs ($\geq 250$). 
	Finally, we removed ASVs which did not map to Bacteria (e.g. Archaeal ASVs) by running Mothur \citep{mothur2009} v1.48.5, a software specialized for 16S sequence analysis.
	After filtering, the result was a collection of 3,851 samples with a collective total of 12,765 ASVs.
	These samples are largely concentrated in the USA, UK and Canada, although several other regions across the globe are represented as well.
	
	\subsection{MetaPhlAn4 study (WGS)}
	
	The data used in \citet{metaphlan4}, itself a compilation of multiple datasets, was requested from the authors of that work. 
	We obtained the abundance table output of MetaPhlAn4 on all samples, as well as the intermediate outputs of PhyloPhlAn3.
	The latter was required, because there are actually two sets of marker genes per SGB: the marker genes used for SGB abundance estimation (MetaPhlAn4 markers, $\sim$200 genes per SGB), and the genes used to construct the phylogenetic tree (PhyloPhlAn3 markers, $\sim$300 genes per SGB). 
	We began with 6,053 samples from healthy adults; we removed the 38 samples which had $\leq 50$ SGBs. 
	The resulting dataset consisted of 6015 samples; the dataset is most largely concentrated in the USA, UK and China, though various countries are represented across the globe.

	\section{Extra Figures and Tables}

\clearpage
\begin{table}[!h]
	\centering
	\begin{tabular}{lcc}
		\toprule
		& JS-Maximizing Split (KL) & Random Split (KL) \\
		\midrule
		Method &  &  \\
		\midrule
		\midrule
		Uniform & 2.6562 & 1.6868 \\
		\midrule
		NN-Global & 1.9337 & 0.8676 \\
		\midrule
		kNN ($k=10$) & 1.8382 & 0.8353 \\
		\midrule
		kNN ($k=20$) & 1.7981 & 0.8089 \\
		\midrule
		kNN ($k=30$) & 1.8044 & 0.7894 \\
		\midrule
		kNN ($k=40$) & 1.8097 & 0.8039 \\
		\midrule
		DNABERT-S & 0.9374 & 0.5705 \\
		\midrule
		Evo 1 & 0.8972 & 0.5444 \\
		\midrule
		Evo 2 & \textbf{0.8578} & \textbf{0.5372} \\
		\midrule
		UMAP ($d=100$) & 0.9528 & 0.6718 \\
		\bottomrule
	\end{tabular}
	\vspace{4pt}
	\caption{(16S ASV dataset) Test set KL divergence evaluation of methods. \textbf{JS-Maximizing Split}: median test KL for the $D_{JS}$ maximizing split. \textbf{Random Split}: median-of-median test KL across ten random seeds. The minimal value in each column is highlighted in bold.}
	\label{table:benchmark_asv}
\end{table}

\begin{table}[!h]
	\centering
	\resizebox{\textwidth}{!}{
		\begin{tabular}{lcccccccccc}
			\toprule
			& Uniform & NN-Global & kNN ($k=10$) & kNN ($k=20$) & kNN ($k=30$) & kNN ($k=40$) & DNABERT-S & Evo 1 & Evo 2 & UMAP ($d=100$) \\
			\midrule
			Uniform & -- & \textbf{0.000}$^*$ & \textbf{0.000}$^*$ & \textbf{0.000}$^*$ & \textbf{0.000}$^*$ & \textbf{0.000}$^*$ & \textbf{0.000}$^*$ & \textbf{0.000}$^*$ & \textbf{0.000}$^*$ & \textbf{0.000}$^*$ \\
			NN-Global & \textbf{0.000}$^*$ & -- & {\color{gray}0.168} & \textbf{0.001}$^*$ & \textbf{0.000}$^*$ & \textbf{0.000}$^*$ & \textbf{0.000}$^*$ & \textbf{0.000}$^*$ & \textbf{0.000}$^*$ & \textbf{0.000}$^*$ \\
			kNN ($k=10$) & \textbf{0.000}$^*$ & {\color{gray}0.168} & -- & \textbf{0.037}$^*$ & \textbf{0.023}$^*$ & {\color{gray}0.053} & \textbf{0.000}$^*$ & \textbf{0.000}$^*$ & \textbf{0.000}$^*$ & \textbf{0.000}$^*$ \\
			kNN ($k=20$) & \textbf{0.000}$^*$ & \textbf{0.001}$^*$ & \textbf{0.037}$^*$ & -- & {\color{gray}0.891} & {\color{gray}0.850} & \textbf{0.000}$^*$ & \textbf{0.000}$^*$ & \textbf{0.000}$^*$ & \textbf{0.000}$^*$ \\
			kNN ($k=30$) & \textbf{0.000}$^*$ & \textbf{0.000}$^*$ & \textbf{0.023}$^*$ & {\color{gray}0.891} & -- & {\color{gray}0.479} & \textbf{0.000}$^*$ & \textbf{0.000}$^*$ & \textbf{0.000}$^*$ & \textbf{0.000}$^*$ \\
			kNN ($k=40$) & \textbf{0.000}$^*$ & \textbf{0.000}$^*$ & {\color{gray}0.053} & {\color{gray}0.850} & {\color{gray}0.479} & -- & \textbf{0.000}$^*$ & \textbf{0.000}$^*$ & \textbf{0.000}$^*$ & \textbf{0.000}$^*$ \\
			DNABERT-S & \textbf{0.000}$^*$ & \textbf{0.000}$^*$ & \textbf{0.000}$^*$ & \textbf{0.000}$^*$ & \textbf{0.000}$^*$ & \textbf{0.000}$^*$ & -- & \textbf{0.000}$^*$ & \textbf{0.000}$^*$ & {\color{gray}0.290} \\
			Evo 1 & \textbf{0.000}$^*$ & \textbf{0.000}$^*$ & \textbf{0.000}$^*$ & \textbf{0.000}$^*$ & \textbf{0.000}$^*$ & \textbf{0.000}$^*$ & \textbf{0.000}$^*$ & -- & \textbf{0.000}$^*$ & \textbf{0.006}$^*$ \\
			Evo 2 & \textbf{0.000}$^*$ & \textbf{0.000}$^*$ & \textbf{0.000}$^*$ & \textbf{0.000}$^*$ & \textbf{0.000}$^*$ & \textbf{0.000}$^*$ & \textbf{0.000}$^*$ & \textbf{0.000}$^*$ & -- & \textbf{0.000}$^*$ \\
			UMAP ($d=100$) & \textbf{0.000}$^*$ & \textbf{0.000}$^*$ & \textbf{0.000}$^*$ & \textbf{0.000}$^*$ & \textbf{0.000}$^*$ & \textbf{0.000}$^*$ & {\color{gray}0.290} & \textbf{0.006}$^*$ & \textbf{0.000}$^*$ & -- \\
			\bottomrule
		\end{tabular}
	}
	\vspace{4pt}
	\caption{(16S ASV dataset) Pairwise Wilcoxon signed-rank, two-tailed p-values (BH-corrected) on per-sample KL, JS-maximizing split. Bold$^*$: $p < 0.05$.}
	\label{table:pval_pcoa_asv}
\end{table}
	
\begin{table}[!h]
	\centering
	\resizebox{\textwidth}{!}{
		\begin{tabular}{lcccccccccc}
			\toprule
			& Uniform & NN-Global & kNN ($k=10$) & kNN ($k=20$) & kNN ($k=30$) & kNN ($k=40$) & DNABERT-S & Evo 1 & Evo 2 & UMAP ($d=100$) \\
			\midrule
			Uniform & -- & \textbf{0.000}$^*$ & \textbf{0.000}$^*$ & \textbf{0.000}$^*$ & \textbf{0.000}$^*$ & \textbf{0.000}$^*$ & \textbf{0.000}$^*$ & \textbf{0.000}$^*$ & \textbf{0.000}$^*$ & \textbf{0.000}$^*$ \\
			NN-Global & \textbf{0.000}$^*$ & -- & \textbf{0.046}$^*$ & \textbf{0.000}$^*$ & \textbf{0.000}$^*$ & \textbf{0.000}$^*$ & \textbf{0.000}$^*$ & \textbf{0.000}$^*$ & \textbf{0.000}$^*$ & \textbf{0.000}$^*$ \\
			kNN ($k=10$) & \textbf{0.000}$^*$ & \textbf{0.046}$^*$ & -- & \textbf{0.000}$^*$ & \textbf{0.000}$^*$ & \textbf{0.000}$^*$ & \textbf{0.000}$^*$ & \textbf{0.000}$^*$ & \textbf{0.000}$^*$ & \textbf{0.000}$^*$ \\
			kNN ($k=20$) & \textbf{0.000}$^*$ & \textbf{0.000}$^*$ & \textbf{0.000}$^*$ & -- & \textbf{0.021}$^*$ & \textbf{0.029}$^*$ & \textbf{0.000}$^*$ & \textbf{0.000}$^*$ & \textbf{0.000}$^*$ & \textbf{0.000}$^*$ \\
			kNN ($k=30$) & \textbf{0.000}$^*$ & \textbf{0.000}$^*$ & \textbf{0.000}$^*$ & \textbf{0.021}$^*$ & -- & {\color{gray}0.185} & \textbf{0.000}$^*$ & \textbf{0.000}$^*$ & \textbf{0.000}$^*$ & \textbf{0.000}$^*$ \\
			kNN ($k=40$) & \textbf{0.000}$^*$ & \textbf{0.000}$^*$ & \textbf{0.000}$^*$ & \textbf{0.029}$^*$ & {\color{gray}0.185} & -- & \textbf{0.000}$^*$ & \textbf{0.000}$^*$ & \textbf{0.000}$^*$ & \textbf{0.000}$^*$ \\
			DNABERT-S & \textbf{0.000}$^*$ & \textbf{0.000}$^*$ & \textbf{0.000}$^*$ & \textbf{0.000}$^*$ & \textbf{0.000}$^*$ & \textbf{0.000}$^*$ & -- & \textbf{0.000}$^*$ & \textbf{0.000}$^*$ & \textbf{0.000}$^*$ \\
			Evo 1 & \textbf{0.000}$^*$ & \textbf{0.000}$^*$ & \textbf{0.000}$^*$ & \textbf{0.000}$^*$ & \textbf{0.000}$^*$ & \textbf{0.000}$^*$ & \textbf{0.000}$^*$ & -- & {\color{gray}0.986} & \textbf{0.000}$^*$ \\
			Evo 2 & \textbf{0.000}$^*$ & \textbf{0.000}$^*$ & \textbf{0.000}$^*$ & \textbf{0.000}$^*$ & \textbf{0.000}$^*$ & \textbf{0.000}$^*$ & \textbf{0.000}$^*$ & {\color{gray}0.986} & -- & \textbf{0.000}$^*$ \\
			UMAP ($d=100$) & \textbf{0.000}$^*$ & \textbf{0.000}$^*$ & \textbf{0.000}$^*$ & \textbf{0.000}$^*$ & \textbf{0.000}$^*$ & \textbf{0.000}$^*$ & \textbf{0.000}$^*$ & \textbf{0.000}$^*$ & \textbf{0.000}$^*$ & -- \\
			\bottomrule
		\end{tabular}
	}
	\vspace{4pt}
	\caption{(16S ASV dataset) Pairwise Wilcoxon signed-rank, two-tailed p-values (BH-corrected) on per-sample KL, random split (single seed). Bold$^*$: $p < 0.05$.}
	\label{table:pval_rng_asv}
\end{table}

	\clearpage
	\begin{table}[!h]
		\centering
		\begin{tabular}{lcc}
			\toprule
			& JS-Maximizing Split (KL) & Random Split (KL) \\
			\midrule
			Method &  &  \\
			\midrule
			\midrule
			Uniform & 1.9565 & 1.6196 \\
			\midrule
			NN-Global & 1.9614 & 0.9535 \\
			\midrule
			kNN ($k=10$) & 1.8006 & 0.8155 \\
			\midrule
			kNN ($k=20$) & 1.6577 & 0.7746 \\
			\midrule
			kNN ($k=30$) & 1.5987 & 0.7623 \\
			\midrule
			kNN ($k=40$) & 1.5917 & 0.7568 \\
			\midrule
			DNABERT-S & 1.1993 & 0.6990 \\
			\midrule
			Evo 1 & 1.1533 & 0.7125 \\
			\midrule
			Evo 2 & \textbf{1.1272} & \textbf{0.6985} \\
			\midrule
			UMAP ($d=100$) & 1.3207 & 0.9378 \\
			\bottomrule
		\end{tabular}
		\vspace{4pt}
		\captionof{table}{
			(WMS SGB dataset) 
			Test set KL divergence evaluation of method.
			\textbf{JS-Maximizing Split}: median test KL for the $D_{JS}$ maximizing split.
			\textbf{Random Split}: median-of-median test KL across ten random seeds.
			The minimal (best) loss value in each column is highlighted in bold.
		}
		\label{table:benchmark_sgb}
	\end{table}
	
	\begin{table}[!h]
		\centering
		\resizebox{\textwidth}{!}{
			\begin{tabular}{lcccccccccc}
				\toprule
				& Uniform & NN-Global & kNN ($k=10$) & kNN ($k=20$) & kNN ($k=30$) & kNN ($k=40$) & DNABERT-S & Evo 1 & Evo 2 & UMAP ($d=100$) \\
				\midrule
				Uniform & -- & \textbf{0.045}$^*$ & \textbf{0.000}$^*$ & \textbf{0.000}$^*$ & \textbf{0.000}$^*$ & \textbf{0.000}$^*$ & \textbf{0.000}$^*$ & \textbf{0.000}$^*$ & \textbf{0.000}$^*$ & \textbf{0.000}$^*$ \\
				NN-Global & \textbf{0.045}$^*$ & -- & \textbf{0.000}$^*$ & \textbf{0.000}$^*$ & \textbf{0.000}$^*$ & \textbf{0.000}$^*$ & \textbf{0.000}$^*$ & \textbf{0.000}$^*$ & \textbf{0.000}$^*$ & \textbf{0.000}$^*$ \\
				kNN ($k=10$) & \textbf{0.000}$^*$ & \textbf{0.000}$^*$ & -- & \textbf{0.000}$^*$ & \textbf{0.000}$^*$ & \textbf{0.000}$^*$ & \textbf{0.000}$^*$ & \textbf{0.000}$^*$ & \textbf{0.000}$^*$ & \textbf{0.000}$^*$ \\
				kNN ($k=20$) & \textbf{0.000}$^*$ & \textbf{0.000}$^*$ & \textbf{0.000}$^*$ & -- & \textbf{0.000}$^*$ & \textbf{0.000}$^*$ & \textbf{0.000}$^*$ & \textbf{0.000}$^*$ & \textbf{0.000}$^*$ & \textbf{0.000}$^*$ \\
				kNN ($k=30$) & \textbf{0.000}$^*$ & \textbf{0.000}$^*$ & \textbf{0.000}$^*$ & \textbf{0.000}$^*$ & -- & \textbf{0.000}$^*$ & \textbf{0.000}$^*$ & \textbf{0.000}$^*$ & \textbf{0.000}$^*$ & \textbf{0.000}$^*$ \\
				kNN ($k=40$) & \textbf{0.000}$^*$ & \textbf{0.000}$^*$ & \textbf{0.000}$^*$ & \textbf{0.000}$^*$ & \textbf{0.000}$^*$ & -- & \textbf{0.000}$^*$ & \textbf{0.000}$^*$ & \textbf{0.000}$^*$ & \textbf{0.000}$^*$ \\
				DNABERT-S & \textbf{0.000}$^*$ & \textbf{0.000}$^*$ & \textbf{0.000}$^*$ & \textbf{0.000}$^*$ & \textbf{0.000}$^*$ & \textbf{0.000}$^*$ & -- & \textbf{0.000}$^*$ & \textbf{0.000}$^*$ & \textbf{0.000}$^*$ \\
				Evo 1 & \textbf{0.000}$^*$ & \textbf{0.000}$^*$ & \textbf{0.000}$^*$ & \textbf{0.000}$^*$ & \textbf{0.000}$^*$ & \textbf{0.000}$^*$ & \textbf{0.000}$^*$ & -- & \textbf{0.000}$^*$ & \textbf{0.000}$^*$ \\
				Evo 2 & \textbf{0.000}$^*$ & \textbf{0.000}$^*$ & \textbf{0.000}$^*$ & \textbf{0.000}$^*$ & \textbf{0.000}$^*$ & \textbf{0.000}$^*$ & \textbf{0.000}$^*$ & \textbf{0.000}$^*$ & -- & \textbf{0.000}$^*$ \\
				UMAP ($d=100$) & \textbf{0.000}$^*$ & \textbf{0.000}$^*$ & \textbf{0.000}$^*$ & \textbf{0.000}$^*$ & \textbf{0.000}$^*$ & \textbf{0.000}$^*$ & \textbf{0.000}$^*$ & \textbf{0.000}$^*$ & \textbf{0.000}$^*$ & -- \\
				\bottomrule
			\end{tabular}
		}
		\vspace{4pt}
		\caption{(WMS SGB dataset)  Pairwise Wilcoxon signed-rank p-values (BH-corrected) on per-sample KL, JS-maximizing split. Bold$^*$: $p < 0.05$.}
		\label{table:pval_pcoa_sgb}
	\end{table}

	\begin{table}[!h]
		\centering
		\resizebox{\textwidth}{!}{
			\begin{tabular}{lcccccccccc}
				\toprule
				& Uniform & NN-Global & kNN ($k=10$) & kNN ($k=20$) & kNN ($k=30$) & kNN ($k=40$) & DNABERT-S & Evo 1 & Evo 2 & UMAP ($d=100$) \\
				\midrule
				Uniform & -- & \textbf{0.000}$^*$ & \textbf{0.000}$^*$ & \textbf{0.000}$^*$ & \textbf{0.000}$^*$ & \textbf{0.000}$^*$ & \textbf{0.000}$^*$ & \textbf{0.000}$^*$ & \textbf{0.000}$^*$ & \textbf{0.000}$^*$ \\
				NN-Global & \textbf{0.000}$^*$ & -- & \textbf{0.000}$^*$ & \textbf{0.000}$^*$ & \textbf{0.000}$^*$ & \textbf{0.000}$^*$ & \textbf{0.000}$^*$ & \textbf{0.000}$^*$ & \textbf{0.000}$^*$ & {\color{gray}0.125} \\
				kNN ($k=10$) & \textbf{0.000}$^*$ & \textbf{0.000}$^*$ & -- & \textbf{0.000}$^*$ & \textbf{0.000}$^*$ & \textbf{0.000}$^*$ & \textbf{0.000}$^*$ & \textbf{0.000}$^*$ & \textbf{0.000}$^*$ & \textbf{0.000}$^*$ \\
				kNN ($k=20$) & \textbf{0.000}$^*$ & \textbf{0.000}$^*$ & \textbf{0.000}$^*$ & -- & \textbf{0.000}$^*$ & \textbf{0.000}$^*$ & \textbf{0.000}$^*$ & \textbf{0.000}$^*$ & \textbf{0.000}$^*$ & \textbf{0.000}$^*$ \\
				kNN ($k=30$) & \textbf{0.000}$^*$ & \textbf{0.000}$^*$ & \textbf{0.000}$^*$ & \textbf{0.000}$^*$ & -- & \textbf{0.000}$^*$ & \textbf{0.000}$^*$ & \textbf{0.000}$^*$ & \textbf{0.000}$^*$ & \textbf{0.000}$^*$ \\
				kNN ($k=40$) & \textbf{0.000}$^*$ & \textbf{0.000}$^*$ & \textbf{0.000}$^*$ & \textbf{0.000}$^*$ & \textbf{0.000}$^*$ & -- & \textbf{0.000}$^*$ & \textbf{0.000}$^*$ & \textbf{0.000}$^*$ & \textbf{0.000}$^*$ \\
				DNABERT-S & \textbf{0.000}$^*$ & \textbf{0.000}$^*$ & \textbf{0.000}$^*$ & \textbf{0.000}$^*$ & \textbf{0.000}$^*$ & \textbf{0.000}$^*$ & -- & \textbf{0.000}$^*$ & \textbf{0.020}$^*$ & \textbf{0.000}$^*$ \\
				Evo 1 & \textbf{0.000}$^*$ & \textbf{0.000}$^*$ & \textbf{0.000}$^*$ & \textbf{0.000}$^*$ & \textbf{0.000}$^*$ & \textbf{0.000}$^*$ & \textbf{0.000}$^*$ & -- & \textbf{0.000}$^*$ & \textbf{0.000}$^*$ \\
				Evo 2 & \textbf{0.000}$^*$ & \textbf{0.000}$^*$ & \textbf{0.000}$^*$ & \textbf{0.000}$^*$ & \textbf{0.000}$^*$ & \textbf{0.000}$^*$ & \textbf{0.020}$^*$ & \textbf{0.000}$^*$ & -- & \textbf{0.000}$^*$ \\
				UMAP ($d=100$) & \textbf{0.000}$^*$ & {\color{gray}0.125} & \textbf{0.000}$^*$ & \textbf{0.000}$^*$ & \textbf{0.000}$^*$ & \textbf{0.000}$^*$ & \textbf{0.000}$^*$ & \textbf{0.000}$^*$ & \textbf{0.000}$^*$ & -- \\
				\bottomrule
			\end{tabular}
		}
		\vspace{4pt}
		\caption{(WMS SGB dataset)  Pairwise Wilcoxon signed-rank p-values (BH-corrected) on per-sample KL, random split (single seed). Bold$^*$: $p < 0.05$.}
		\label{table:pval_rng_sgb}
	\end{table}

	\clearpage
	\begin{figure}[h]
		\begin{center}
			\centerline{\includegraphics[width=\columnwidth]{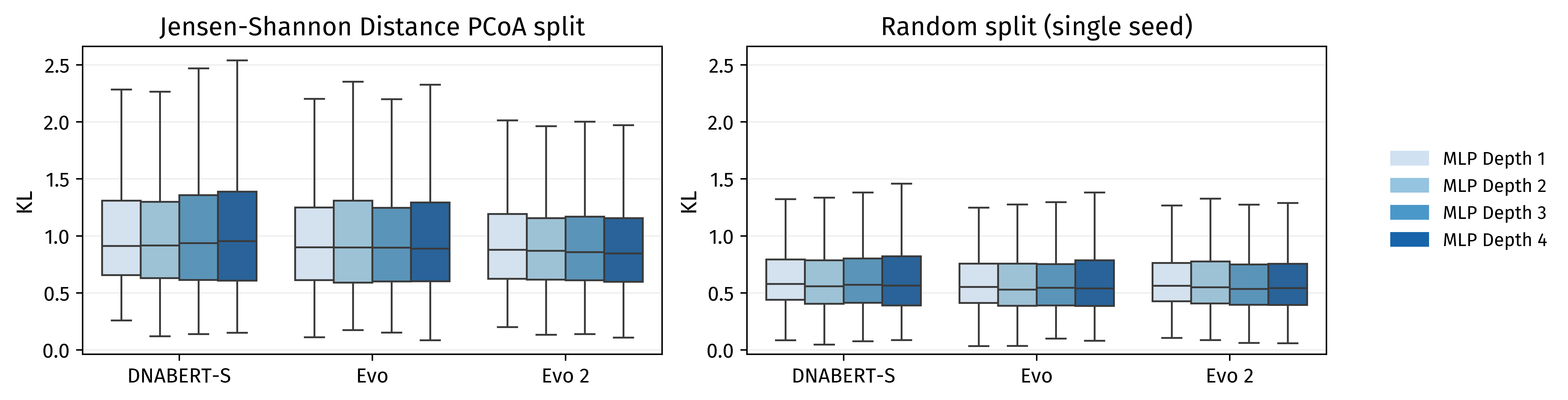}}
			\centerline{\includegraphics[width=\columnwidth]{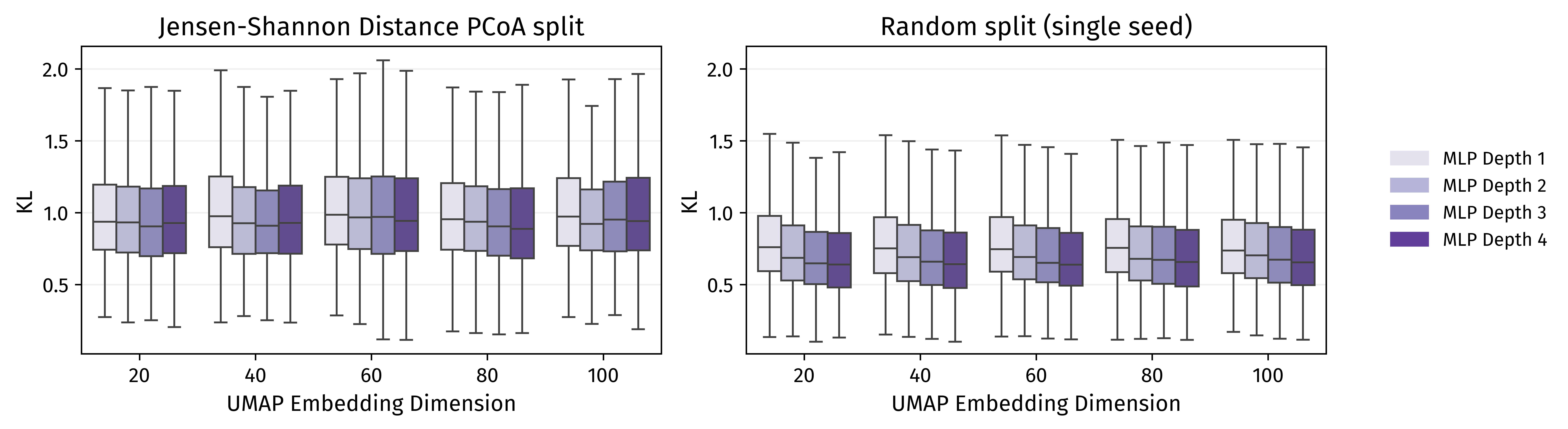}}
			\caption{
				Model ablation comparison, globally varying the MLP capacity by depth. Each hidden layer's width is kept constant (width=32). The sequence embedding-based models are already saturated in performance and do not improve with increased depth. The same is true of UMAP but only for the PCoA split; it consistently improves in performance across all UMAP output dimensions as MLP depth increases.
			}
			\label{fig:mlp_depths}
		\end{center}
	\end{figure}

	\begin{figure}[h]
		\begin{center}
			\centerline{\includegraphics[width=\columnwidth]{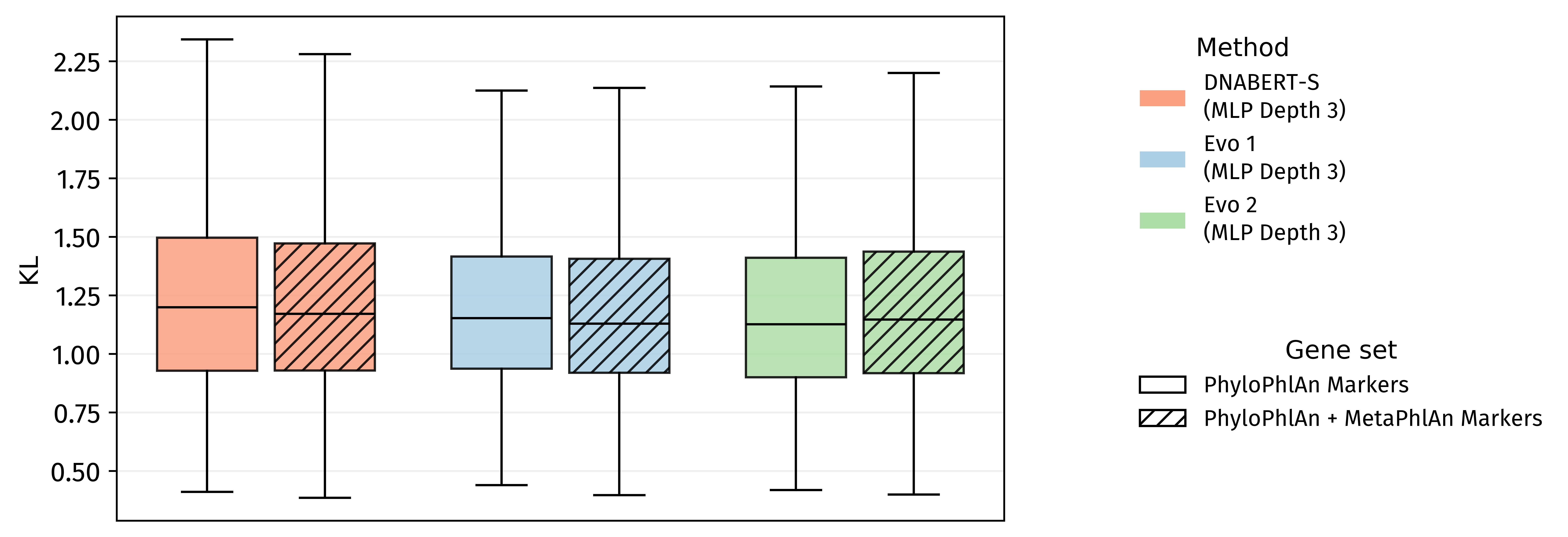}}
			\caption{
				A feature-ablation analysis where we train two families of models: one that uses PhyloPhlAn SGB markers (found in the main text), and another that uses both PhyloPhlAn SGB markers and MetaPhlAn4 SGB markers.
			}
			\label{fig:metaphlan_gene_set}
		\end{center}
	\end{figure}

	
\end{document}